\documentclass[prl,aps,twocolumn,superscriptaddress,showpacs,preprintnumbers,amsmath,amssymb,nofootinbib]{revtex4-1} 
\usepackage{graphicx}
\usepackage{xspace}
\usepackage{epsfig}
\usepackage{amsfonts}
\usepackage{multirow}
\usepackage{dcolumn}  
\usepackage{wasysym}
\usepackage[section]{placeins}
\usepackage{multirow}
\usepackage{subfigure}

\usepackage[colorlinks,linkcolor=black,citecolor=black]{hyperref}

\newcommand{\mb}{\ensuremath{M_{\mathrm{bc}}}\xspace}
\newcommand{\de}{\ensuremath{\Delta E}\xspace}

\newcommand{\acp}{\ensuremath{\mathcal{A}_{\mathit CP}}\xspace}
\newcommand{\br}{\ensuremath{\mathcal{B}}\xspace}

\newcommand{\ks}{\ensuremath{K^0_S}\xspace}
\newcommand{\bb}{\ensuremath{B \bar{B}}\xspace}
\newcommand{\mkk}{\ensuremath{M_{K^+K^-}}\xspace}
\newcommand{\kk}{\ensuremath{{K^+K^-}}\xspace}
\newcommand{\kpi}{\ensuremath{{K^+ \pi^-}}\xspace}
\newcommand{\ebeam}{E_\mathrm{beam}}

\newcommand{\Nsig}{N_{\rm sig}}
\newcommand{\kkpi}{\ensuremath{{B^+ \to K^+K^- \pi^+}}\xspace}
\newcommand{\mkkpi}{\ensuremath{{B^- \to K^-K^+ \pi^-}}\xspace}
\newcommand{\control}{\ensuremath{{B^+ \to \bar{D}^0(K^{+}K^{-}) \pi^+}}\xspace}

\def\myspecial#1{}                   

\def\BR{{\mathcal B}}
\def\Mbc{M_{\rm bc}}
\def\nn{{\mathcal C_{\mathit NN}}}
\def\cpid{\mathcal{C}_{\mathit PID}}
\def\rkpi{\mathcal{R}_{K/\pi}}

\def\figurebox#1#2#3{%
    \def\arg{#3}%
    \ifx\arg\empty
    {\hfill\vbox{\hsize#2\hrule\hbox to #2{\vrule\hfill\vbox to #1{\hsize#2\vfill}\vrule}\hrule}\hfill}%
    \else
    {\hfill\epsfbox{#3}\hfill}%
    \fi}

\begin{document}
\preprint{\vbox{
							   \hbox{Belle Preprint 2017-09}
							   \hbox{KEK Preprint 2017-5}
}}

\title{
{\Large \bf \boldmath
Measurement of branching fraction and direct $CP$ asymmetry in charmless $\kkpi$ decays at Belle}}

\noaffiliation
\affiliation{University of the Basque Country UPV/EHU, 48080 Bilbao}
\affiliation{University of Bonn, 53115 Bonn}
\affiliation{Budker Institute of Nuclear Physics SB RAS, Novosibirsk 630090}
\affiliation{Faculty of Mathematics and Physics, Charles University, 121 16 Prague}
\affiliation{Chonnam National University, Kwangju 660-701}
\affiliation{University of Cincinnati, Cincinnati, Ohio 45221}
\affiliation{Deutsches Elektronen--Synchrotron, 22607 Hamburg}
\affiliation{University of Florida, Gainesville, Florida 32611}
\affiliation{Justus-Liebig-Universit\"at Gie\ss{}en, 35392 Gie\ss{}en}
\affiliation{SOKENDAI (The Graduate University for Advanced Studies), Hayama 240-0193}
\affiliation{Hanyang University, Seoul 133-791}
\affiliation{University of Hawaii, Honolulu, Hawaii 96822}
\affiliation{High Energy Accelerator Research Organization (KEK), Tsukuba 305-0801}
\affiliation{J-PARC Branch, KEK Theory Center, High Energy Accelerator Research Organization (KEK), Tsukuba 305-0801}
\affiliation{IKERBASQUE, Basque Foundation for Science, 48013 Bilbao}
\affiliation{Indian Institute of Science Education and Research Mohali, SAS Nagar, 140306}
\affiliation{Indian Institute of Technology Bhubaneswar, Satya Nagar 751007}
\affiliation{Indian Institute of Technology Guwahati, Assam 781039}
\affiliation{Indian Institute of Technology Madras, Chennai 600036}
\affiliation{Indiana University, Bloomington, Indiana 47408}
\affiliation{Institute of High Energy Physics, Vienna 1050}
\affiliation{INFN - Sezione di Torino, 10125 Torino}
\affiliation{J. Stefan Institute, 1000 Ljubljana}
\affiliation{Kanagawa University, Yokohama 221-8686}
\affiliation{Institut f\"ur Experimentelle Kernphysik, Karlsruher Institut f\"ur Technologie, 76131 Karlsruhe}
\affiliation{Kennesaw State University, Kennesaw, Georgia 30144}
\affiliation{King Abdulaziz City for Science and Technology, Riyadh 11442}
\affiliation{Department of Physics, Faculty of Science, King Abdulaziz University, Jeddah 21589}
\affiliation{Korea Institute of Science and Technology Information, Daejeon 305-806}
\affiliation{Korea University, Seoul 136-713}
\affiliation{Kyungpook National University, Daegu 702-701}
\affiliation{\'Ecole Polytechnique F\'ed\'erale de Lausanne (EPFL), Lausanne 1015}
\affiliation{P.N. Lebedev Physical Institute of the Russian Academy of Sciences, Moscow 119991}
\affiliation{Faculty of Mathematics and Physics, University of Ljubljana, 1000 Ljubljana}
\affiliation{Ludwig Maximilians University, 80539 Munich}
\affiliation{Luther College, Decorah, Iowa 52101}
\affiliation{University of Maribor, 2000 Maribor}
\affiliation{Max-Planck-Institut f\"ur Physik, 80805 M\"unchen}
\affiliation{School of Physics, University of Melbourne, Victoria 3010}
\affiliation{University of Miyazaki, Miyazaki 889-2192}
\affiliation{Moscow Physical Engineering Institute, Moscow 115409}
\affiliation{Moscow Institute of Physics and Technology, Moscow Region 141700}
\affiliation{Graduate School of Science, Nagoya University, Nagoya 464-8602}
\affiliation{Nara Women's University, Nara 630-8506}
\affiliation{National Central University, Chung-li 32054}
\affiliation{National United University, Miao Li 36003}
\affiliation{Department of Physics, National Taiwan University, Taipei 10617}
\affiliation{H. Niewodniczanski Institute of Nuclear Physics, Krakow 31-342}
\affiliation{Nippon Dental University, Niigata 951-8580}
\affiliation{Niigata University, Niigata 950-2181}
\affiliation{Novosibirsk State University, Novosibirsk 630090}
\affiliation{Osaka City University, Osaka 558-8585}
\affiliation{Pacific Northwest National Laboratory, Richland, Washington 99352}
\affiliation{University of Pittsburgh, Pittsburgh, Pennsylvania 15260}
\affiliation{Theoretical Research Division, Nishina Center, RIKEN, Saitama 351-0198}
\affiliation{University of Science and Technology of China, Hefei 230026}
\affiliation{Showa Pharmaceutical University, Tokyo 194-8543}
\affiliation{Soongsil University, Seoul 156-743}
\affiliation{University of South Carolina, Columbia, South Carolina 29208}
\affiliation{Stefan Meyer Institute for Subatomic Physics, Vienna 1090}
\affiliation{Sungkyunkwan University, Suwon 440-746}
\affiliation{School of Physics, University of Sydney, New South Wales 2006}
\affiliation{Department of Physics, Faculty of Science, University of Tabuk, Tabuk 71451}
\affiliation{Tata Institute of Fundamental Research, Mumbai 400005}
\affiliation{Excellence Cluster Universe, Technische Universit\"at M\"unchen, 85748 Garching}
\affiliation{Department of Physics, Technische Universit\"at M\"unchen, 85748 Garching}
\affiliation{Toho University, Funabashi 274-8510}
\affiliation{Department of Physics, Tohoku University, Sendai 980-8578}
\affiliation{Earthquake Research Institute, University of Tokyo, Tokyo 113-0032}
\affiliation{Department of Physics, University of Tokyo, Tokyo 113-0033}
\affiliation{Tokyo Institute of Technology, Tokyo 152-8550}
\affiliation{Virginia Polytechnic Institute and State University, Blacksburg, Virginia 24061}
\affiliation{Wayne State University, Detroit, Michigan 48202}
\affiliation{Yamagata University, Yamagata 990-8560}
\affiliation{Yonsei University, Seoul 120-749}

  \author{C.-L.~Hsu}\affiliation{School of Physics, University of Melbourne, Victoria 3010} 
  \author{D.~Dossett}\affiliation{School of Physics, University of Melbourne, Victoria 3010} 
  \author{M.~E.~Sevior}\affiliation{School of Physics, University of Melbourne, Victoria 3010} 

  \author{I.~Adachi}\affiliation{High Energy Accelerator Research Organization (KEK), Tsukuba 305-0801}\affiliation{SOKENDAI (The Graduate University for Advanced Studies), Hayama 240-0193} 
  \author{H.~Aihara}\affiliation{Department of Physics, University of Tokyo, Tokyo 113-0033} 
  \author{S.~Al~Said}\affiliation{Department of Physics, Faculty of Science, University of Tabuk, Tabuk 71451}\affiliation{Department of Physics, Faculty of Science, King Abdulaziz University, Jeddah 21589} 
  \author{D.~M.~Asner}\affiliation{Pacific Northwest National Laboratory, Richland, Washington 99352} 
  \author{H.~Atmacan}\affiliation{University of South Carolina, Columbia, South Carolina 29208} 
  \author{V.~Aulchenko}\affiliation{Budker Institute of Nuclear Physics SB RAS, Novosibirsk 630090}\affiliation{Novosibirsk State University, Novosibirsk 630090} 
  \author{T.~Aushev}\affiliation{Moscow Institute of Physics and Technology, Moscow Region 141700} 
% ...
\collaboration{The Belle Collaboration}

\begin{abstract}
We report a study of the charmless hadronic decay of the charged $B$ meson to the three-body final state $K^+ K^- \pi^+$.
The results are based on a data sample that contains $772\times10^6$ $\bb$ pairs collected at the $\Upsilon(4S)$ resonance 
with the Belle detector at the KEKB asymmetric-energy $e^+ e^-$ collider. The measured inclusive branching fraction 
and direct $CP$ asymmetry are $(5.38\pm0.40\pm0.35)\times 10^{-6}$ and $-0.170\pm0.073\pm0.017$, respectively, 
where the first uncertainties are statistical and the second are systematic. 
The $\kk$ invariant mass distribution of the signal candidates shows an excess in the region below $1.5$ GeV/$c^2$, which is consistent with 
the previous studies from $BABAR$ and LHCb. In addition, strong evidence of a large direct $CP$ asymmetry is found in the low $\kk$ invariant-mass region.

\pacs{14.40.Nd,13.25.Hw,11.30.Er}
\end{abstract}

\maketitle
\begin{figure}[htb]
\includegraphics[width=0.23\textwidth]{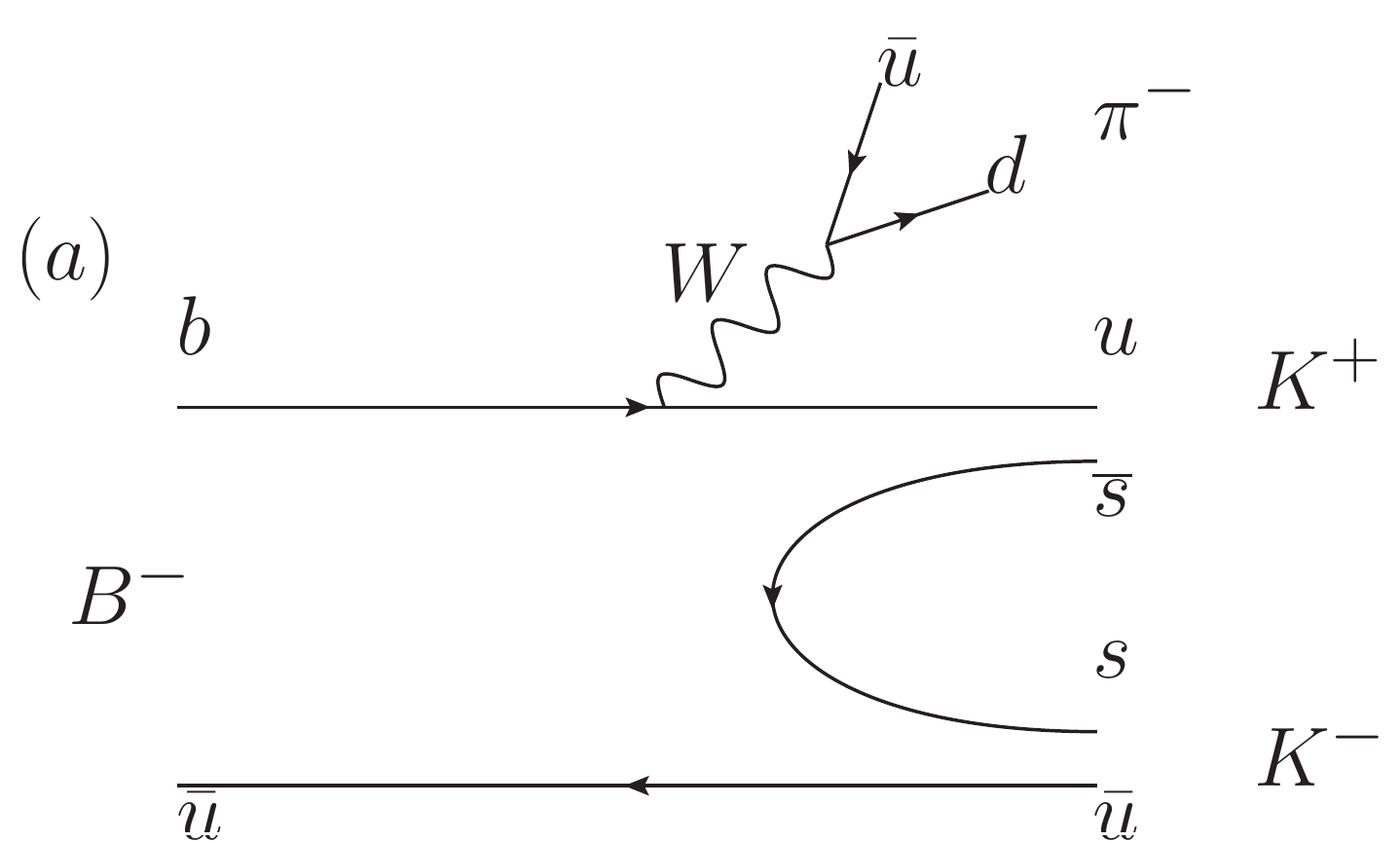} 
\includegraphics[width=0.23\textwidth]{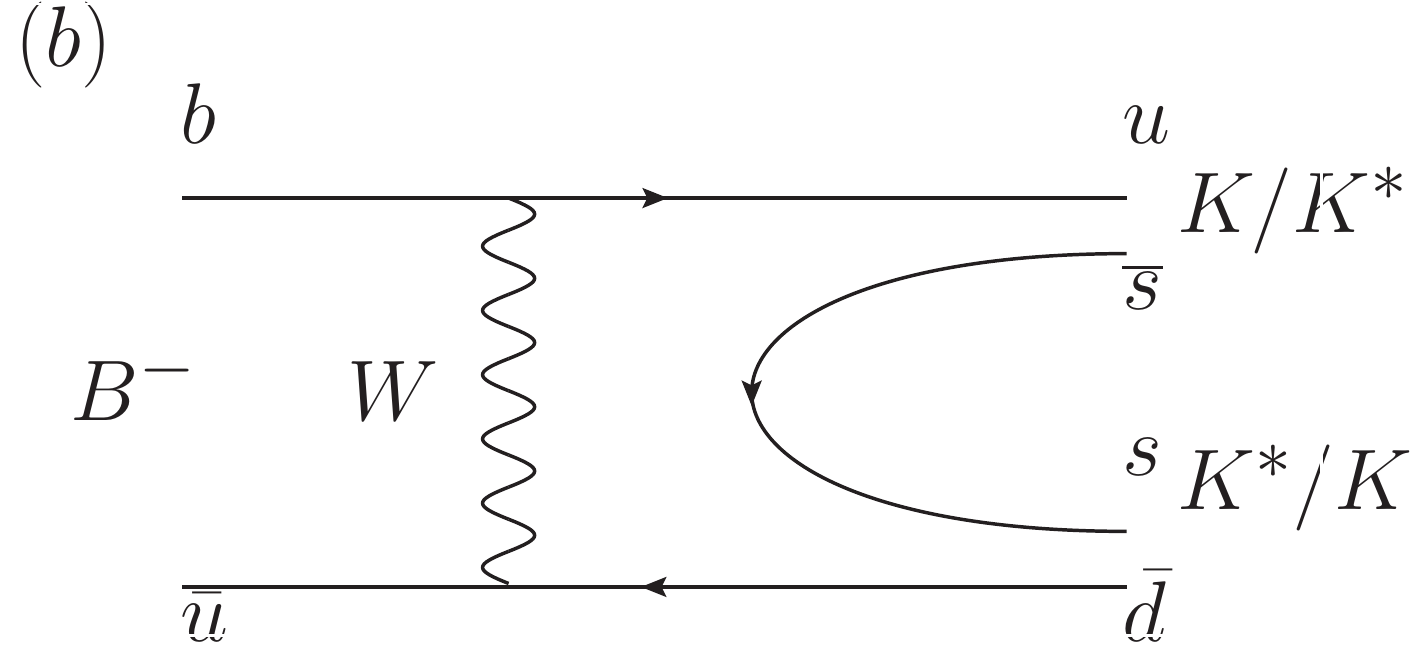} \\
\includegraphics[width=0.23\textwidth]{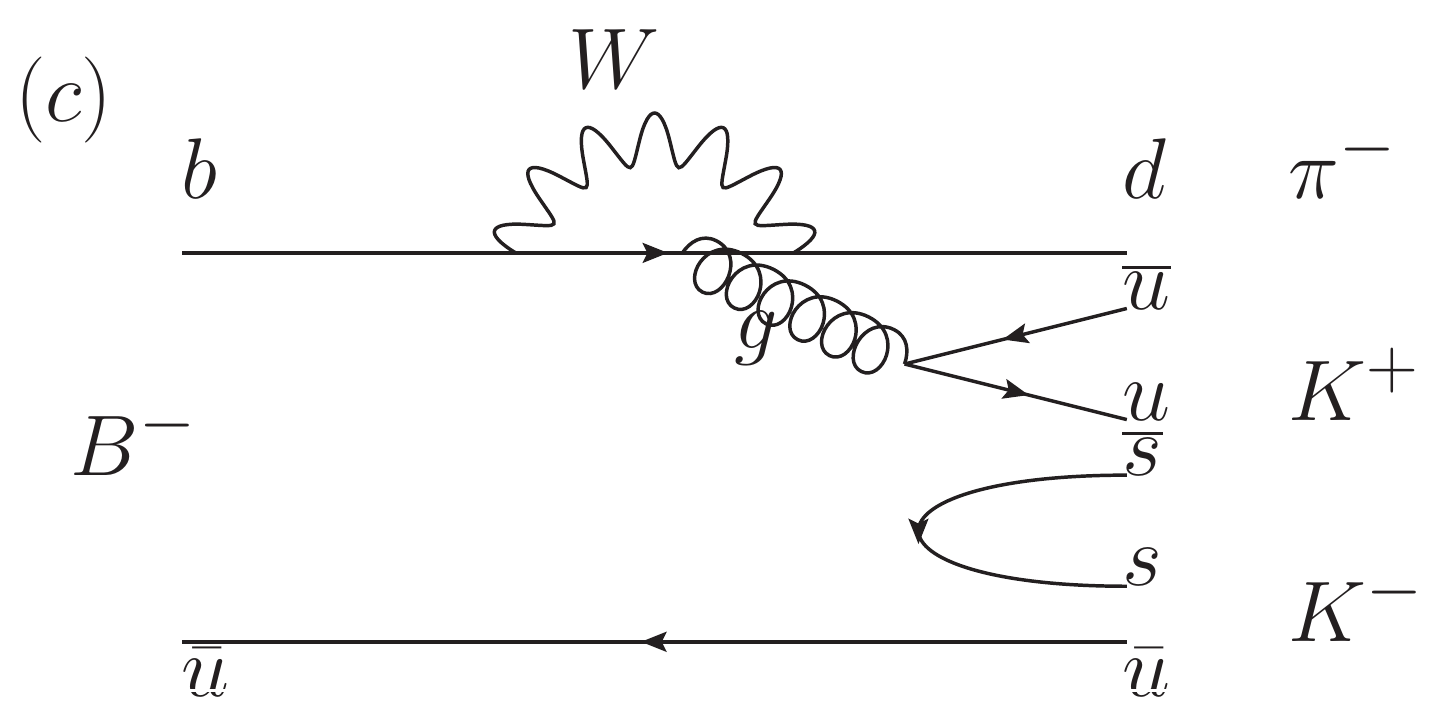}
\includegraphics[width=0.23\textwidth]{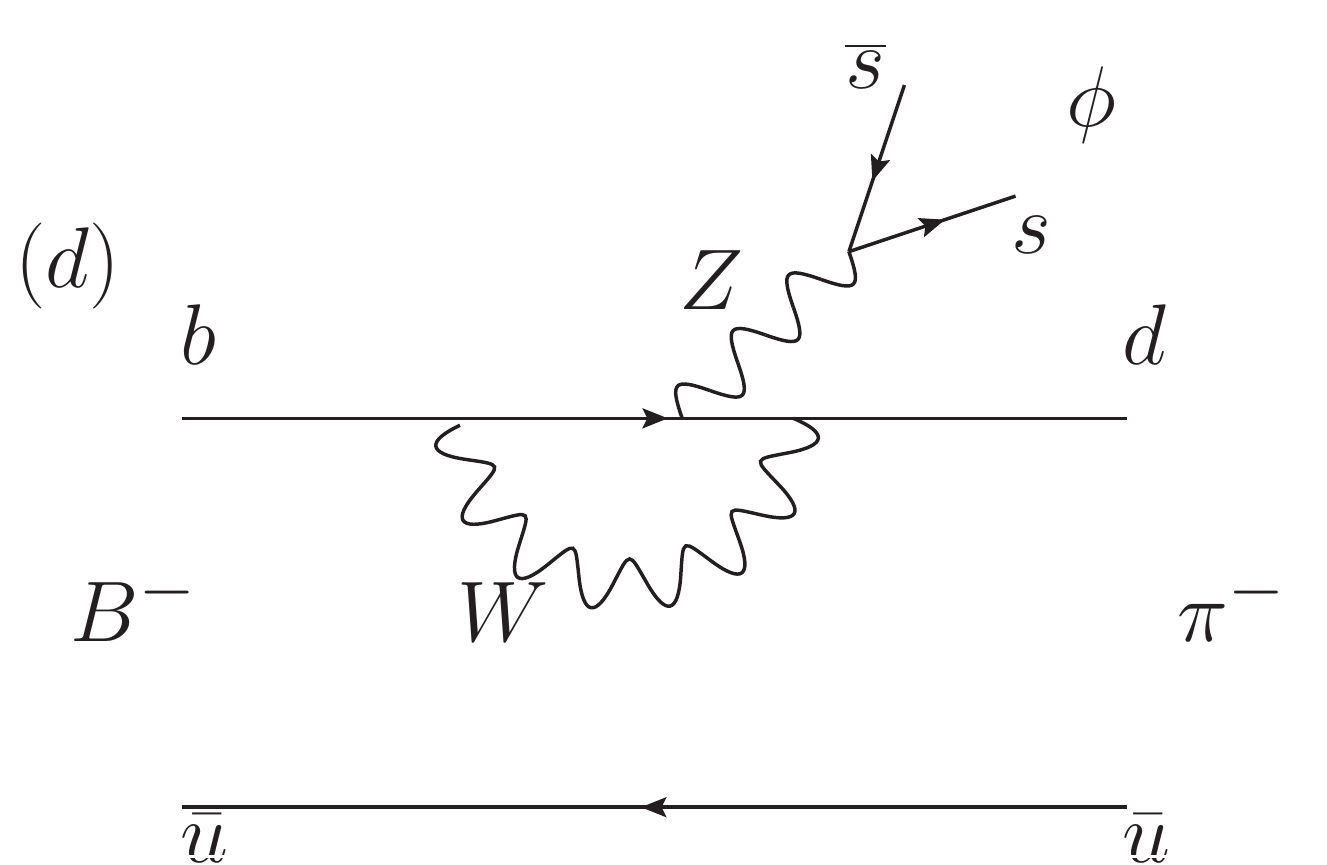} \\  
\caption{$\kkpi$ Feynman diagrams (all Cabibbo-suppressed). (a) Tree diagram, (b) $W$-exchange diagram leading to $KK^{*}$ states, (c) strong-penguin diagram, and (d) electroweak penguin leading to the $\phi\pi$ state.}
\label{fig:fey_dia}
\end{figure}

Charmless decays of $B$ mesons to three charged hadrons are suppressed in the standard model (SM), and thus provide an opportunity to search for physics beyond the SM through branching fraction enhancements.
Large $CP$ asymmetries can occur in these decays, due to interference of tree and loop diagrams with similar amplitudes. Beyond-the-SM particles could also contribute in the loops. 
Figure~\ref{fig:fey_dia} shows 
some of the SM Feynman diagrams that contribute to the $\kkpi$ decay~\footnote{The inclusion of charge-conjugate modes is implied throughout this paper, unless explicitly stated otherwise.}.
The dominant process is the Cabibbo-suppressed $b \to u$ tree transition in Fig~\ref{fig:fey_dia}(a); 
the $b \to d$ penguin diagram in Fig.~\ref{fig:fey_dia}(d) leading to $B^+ \to \phi \pi^{+}$ with $\phi \to \kk$ is heavily suppressed.

In recent years, an unidentified structure has been observed by $BABAR$~\cite{BaBar_kkpi} and LHCb~\cite{lhcb_old,lhcb_new} in the low
$\kk$ invariant-mass spectrum of the $\kkpi$ decay. 
The LHCb studies revealed a nonzero inclusive $CP$ asymmetry of $-0.123\pm0.017\pm0.012\pm0.007$ and a large unquantified local $CP$ asymmetry in the same mass region.
These results suggest that final-state interactions may contribute to $CP$ violation~\cite{Bhattacharya2013,Bediaga2014}.
This study attempts to quantify the $CP$ asymmetry and branching fraction as a function of the $\kk$ invariant mass.

In this paper, we report the measurements of branching fraction and direct $CP$ asymmetry~(\acp) of the charmless $\kkpi$ decay based on the data 
collected with the Belle detector at the KEKB asymmetric-energy (3.5 on 8.0 GeV) 
$e^+e^-$ collider~\cite{kekb}. 
The latter is defined as
\begin{equation}
\label{eq:acpdef}
\acp = \frac{N(\mkkpi)-N(\kkpi)}{N(\mkkpi)+N(\kkpi)}\mbox{,}
\end{equation}
where $N$ denotes the yield obtained for the corresponding mode. 
The data sample consists of $772 \times 10^{6}$ $B\overline{B}$ pairs accumulated at the $\Upsilon(4S)$ resonance, corresponding to an integrated luminosity of $711~\rm fb^{-1}$, and an additional 
$89.4~{\rm fb}^{-1}$ of off resonance data recorded at a center-of-mass (c.m.) energy about $60$ MeV below the $\Upsilon(4S)$ resonance.

The Belle detector consists of a silicon vertex detector (SVD), a 50-layer central drift chamber (CDC), time-of-flight scintillation counters (TOF) an array of aerogel threshold Cherenkov counters (ACC), and a CsI(Tl) electromagnetic calorimeter located inside a superconducting solenoid coil that provides a 1.5~T magnetic field.
Outside the coil, the $K_L^0$ and muon detector, composed of resistive plate counters, detects $K_L^0$ mesons and identifies muons. The detector is described in detail elsewhere~\cite{belled}.
The data set used in this analysis was collected with two different inner detector configurations. A data sample corresponding to $140~\rm fb^{-1}$ was collected with a beam pipe of radius $2.0$ cm and with three layers of SVD, while the rest of the data were collected with a beam pipe of radius 1.5 cm and four layers of SVD~\cite{bellesvd2}.
A GEANT3-based~\cite{geant} Monte Carlo (MC) simulation of the Belle detector is used to optimize event selection and to estimate the signal efficiency.
The signal MC sample is generated with the EvtGen package~\cite{evtgen}, assuming a three-body phase space combined with 
an intermediate resonance decaying to two kaons as observed by $BABAR$ and LHCb~\cite{BaBar_kkpi,lhcb_new}.

To reconstruct $\kkpi$, we combine two oppositely charged kaons with a charged pion. 
Charged tracks originating from a $B$-meson decay are required to have a distance of closest approach with respect to the interaction point of less than 5.0 cm along the $z$ axis (opposite to the positron beam direction) and less than 0.2 cm in the $r$-$\phi$ transverse plane, and a transverse momentum of at least 100 MeV/$c$.

Charged kaons and pions are identified using specific ionization in the CDC, time-of-flight information from the TOF, and Cherenkov light yield in the ACC.
This information is combined to form a $K$-$\pi$ likelihood ratio 
$\mathcal{R}_{K/\pi} = \mathcal{L}_K/(\mathcal{L}_K+\mathcal{L}_\pi)$, 
where ${\cal L}_K$ and ${\cal L}_{\pi}$ are the likelihoods for the kaon and pion hypothesis, respectively. 
Tracks with ${\cal R}_{K/\pi} > 0.6$ are regarded as kaons and those with ${\cal R}_{K/\pi} < 0.4$ as pions. 
With these requirements, the identification efficiencies for 1 GeV/$c$ kaons and pions are $83\%$ and $90\%$, respectively; $6\%$ of 
the pions are misidentified as kaons and $12\%$ of the kaons as pions.

Candidate $B$ mesons are identified using two kinematic variables: the beam-energy constrained mass, $\Mbc \equiv \sqrt{\ebeam^{2}/c^{4}-{|p_{B}/c|}^{2}}$, and the energy difference, $\de \equiv E_{B} - \ebeam$, where $E_{B}$ and $p_{B}$ are the reconstructed energy and momentum of $B$-meson candidates in the c.m. frame, respectively, and $\ebeam$ is the run-dependent beam energy. The fit region is defined as $\mb>5.24$~GeV/$c^2$ and $|\de|<0.3$~GeV, while the signal-enhanced region is given by $5.27$~GeV/$c^2 < \mb < 5.29$~GeV/$c^2$ and $|\de|<0.05$~GeV.
For $19\%$ of the events, there is more than one $B$-meson candidate; we choose the one with the best fit quality from the $B$ vertex fit. The $B$ vertex fit uses the three charged tracks. This criterion selects the correct $B$-meson candidate in $92\%$ of MC events.

The dominant background is from continuum $e^+ e^- \to q\overline{q}~(q=u,d,s,c)$ processes. 
A neural network~\cite{neubay} is used to distinguish the spherical $\bb$ signal from the jetlike continuum background. 
It combines the following observables based on the event topology: a Fisher discriminant formed from 17 modified Fox-Wolfram moments~\cite{ksfw}, 
the cosine of the angle between the $B$-meson candidate direction and the beam axis, 
the cosine of the angle between the thrust axis~\cite{thrust} of the $B$-meson candidate and that of the rest of event 
(all of these quantities being calculated in the c.m. frame), 
the separation along the $z$ axis between the vertex of the $B$-meson candidate and that of the remaining tracks, 
and the tagging quality variable from a $B$-meson flavor-tagging algorithm~\cite{btagging}. 
The training and optimization of the neural network are performed with signal and continuum MC samples. 
The neural network output ($\nn$) selection requirement is optimized by maximizing a figure of merit 
defined as $N_{\textrm S}/\sqrt{N_{\textrm S}+N_{\textrm B}}$ in the signal-enhanced region, where $N_{\textrm S}$ denotes 
the expected number of signal events based on MC simulation for a branching fraction of $5\times 10^{-6}$ 
and $N_{\textrm B}$ denotes the expected number of background events. 
The requirement on $\nn$ removes $99\%$ of the continuum events while retaining $48\%$ of the signal.

Background contributions from $B$-meson decays via the dominant $b \to c$ transition (generic $B$ decays) are investigated with an MC 
sample of such decays. The resulting $\de$ distribution is found to peak strongly in the signal region. 
Peaks are observed in the $\kk$ and $\kpi$ invariant-mass spectra, arising from $b \to c$ decays. These contributions are from $D^{0} \to \kk$ or $K^- \pi^+$ peaking
at the nominal $D^0$ mass, and $D^0\to K^- \pi^+$ with the peak slightly shifted from the $D^0$ mass in the $\mkk$ spectrum owing to $K$-$\pi$ misidentification. 
To suppress these backgrounds, the candidates for which the invariant mass of the $\kk$ or $\kpi$ system lies in range of $1850$-$1880$ MeV/$c^2$ are removed. 
The selection window corresponds to $\pm 3.75\sigma$ around the nominal $D^0$ mass, where $\sigma$ is the mass resolution. 
In the case of $K$-$\pi$ misidentification, we use the pion hypothesis for one of the kaons. 
To suppress the possible charmonium backgrounds from $J/\psi \to \ell^+ \ell^- (\ell=e, \mu)$ decays, 
we apply the lepton hypothesis for both charged daughters and exclude candidates that lie in the range of $3060$-$3140$ MeV/$c^2$, 
which corresponds to $\pm 4\sigma$ around the nominal $J/\psi$ mass. Since no significant resonance is found in the $\psi(2S)$ mass region, we do not apply a veto selection for it.

The charmless (i.e., ``rare'') $B$-meson decays
are studied with a large MC sample where one of the $B$ mesons decays via a process with known or estimated branching fraction.
There are a few modes that contribute in the $\mb$ signal region with a corresponding $\de$ peak, denoted collectively as the ``rare peaking''
background. These peaking backgrounds are due to $K$-$\pi$ misidentification, including $B^+\to \kk K^+$, $B^+\to K^+ \pi^- \pi^+$, 
and their intermediate resonant modes. The events that remain after removing the peaking components are called the ``rare combinatorial'' background.

The signal yield and direct $CP$ asymmetry are extracted by performing a two-dimensional extended unbinned maximum likelihood fit to $\mb$ and $\de$ distributions in bins of $\mkk$. 
In order to study the finer structure in the enhancement region, the $\mkk$ region is divided into five nonuniform bins. The first two bins are chosen to cover the interesting enhancement and $\acp$ signal found by LHCb and $BABAR$. The remaining bin ranges are chosen in order to have an approximately equal number of signal events in each bin. The likelihood is defined as
\begin{equation}
\mathcal{L}=\frac{e^{-\sum_jN_j}}{N!}\prod^{N}_{i}\left(\sum_j N_j\mathcal{P}^{i}_{j}\right)\mbox{,}
\end{equation}
where
\begin{equation}
\mathcal{P}^{i}_{j}=\frac{1}{2}(1-q^{i}\cdot \mathcal{A}_{CP,j}\xspace) \times \mathcal{P}_{j}(\mb^i,\de^i)\mbox{.}
\end{equation}
Here, $N$ is the total number of candidate events, $i$ is the event index, and $N_j$ is the yield of events for category $j$, 
which indexes the signal, continuum, generic $B$, and rare $B$ components. 
$\mathcal{P}_{j}(\mb,\de)$ is the probability density function (PDF) in $\mb$ and $\de$ for the $j{\rm th}$
 category. 
The electric charge of the $B$-meson candidate in event $i$ is denoted $q^{i}$ and $\mathcal{A}_{CP,j}$ is the direct $CP$ asymmetry for category $j$. 
In the signal $B$ decays, there are two cases: all final state particles are correctly combined (``true'' signal), or one of the daughter particles is 
a product of the other $B$-meson decay (``self-cross-feed'' [SCF] background). We prepare the corresponding PDFs, ${\cal P}_{\rm sig}$ and ${\cal P}_{\rm SCF}$. The SCF background is described by $(\Nsig\cdot f) \times \cal{P}_{\rm SCF}$, where $\Nsig$ is the signal yield and $f$ is the fraction of SCF component, which is fixed to the MC expectation.
The signal PDF is represented by the product of a double Gaussian in $\mb$ and a triple Gaussian in $\de$, where the shape parameters are determined from the signal MC sample and are calibrated by a control sample of $\control$.
The PDF that describes the continuum background is the product of an ARGUS function~\cite{argus} in $\mb$ and a second-order polynomial in $\de$.
The parameters of the continuum PDF are derived from MC simulation, which agree with the off resonance data.
In contrast, the distributions for $\de$ and $\mb$ are strongly correlated in the $\bb$ background samples, including generic $B$, 
rare combinatorial, rare peaking, and SCF components. 
Therefore, they are modeled with two-dimensional smoothed histograms from MC simulation. 
The free parameters in the fit are the signal yield, the signal $\acp$, the generic $B$ yield, the rare peaking yields, and the continuum yield. 
The yields of rare combinatorial backgrounds are also derived from the MC study. 
The $\acp$ of all backgrounds is fixed to zero in the fit.
The stability and bias of the two-dimensional fit is checked by large ensemble tests using both toy and simulated MC events. 
The validity of the fit and branching fraction extraction method is checked using data in a high-statistics control sample of the $\control$ decays.
The measured branching fraction for the control sample is consistent with the world-average value~\cite{pdg2016}. 

\begin{figure}
 \subfigure[$\kkpi$]{  \includegraphics[width=0.48\textwidth]{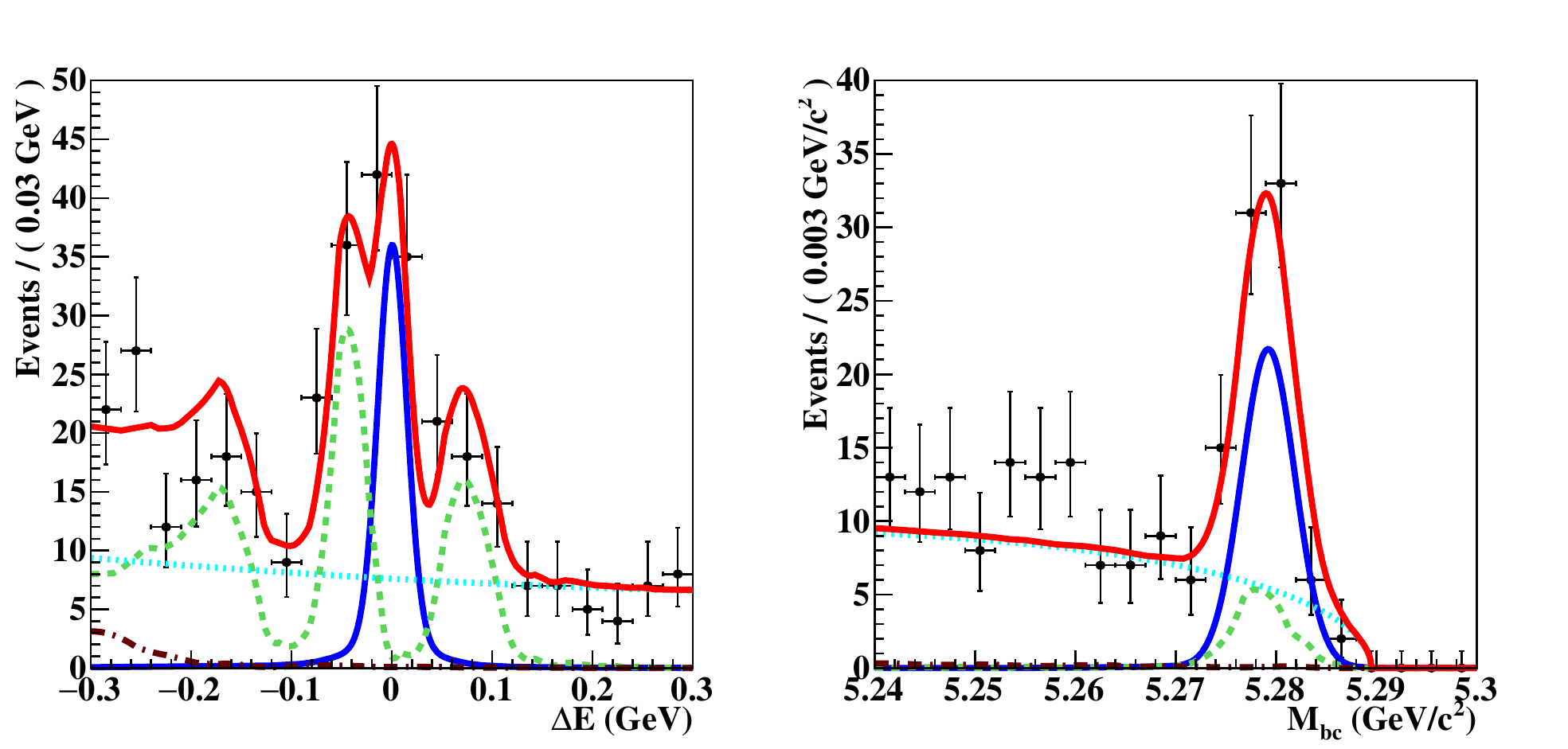}
}
 \subfigure[$\mkkpi$]{ \includegraphics[width=0.48\textwidth]{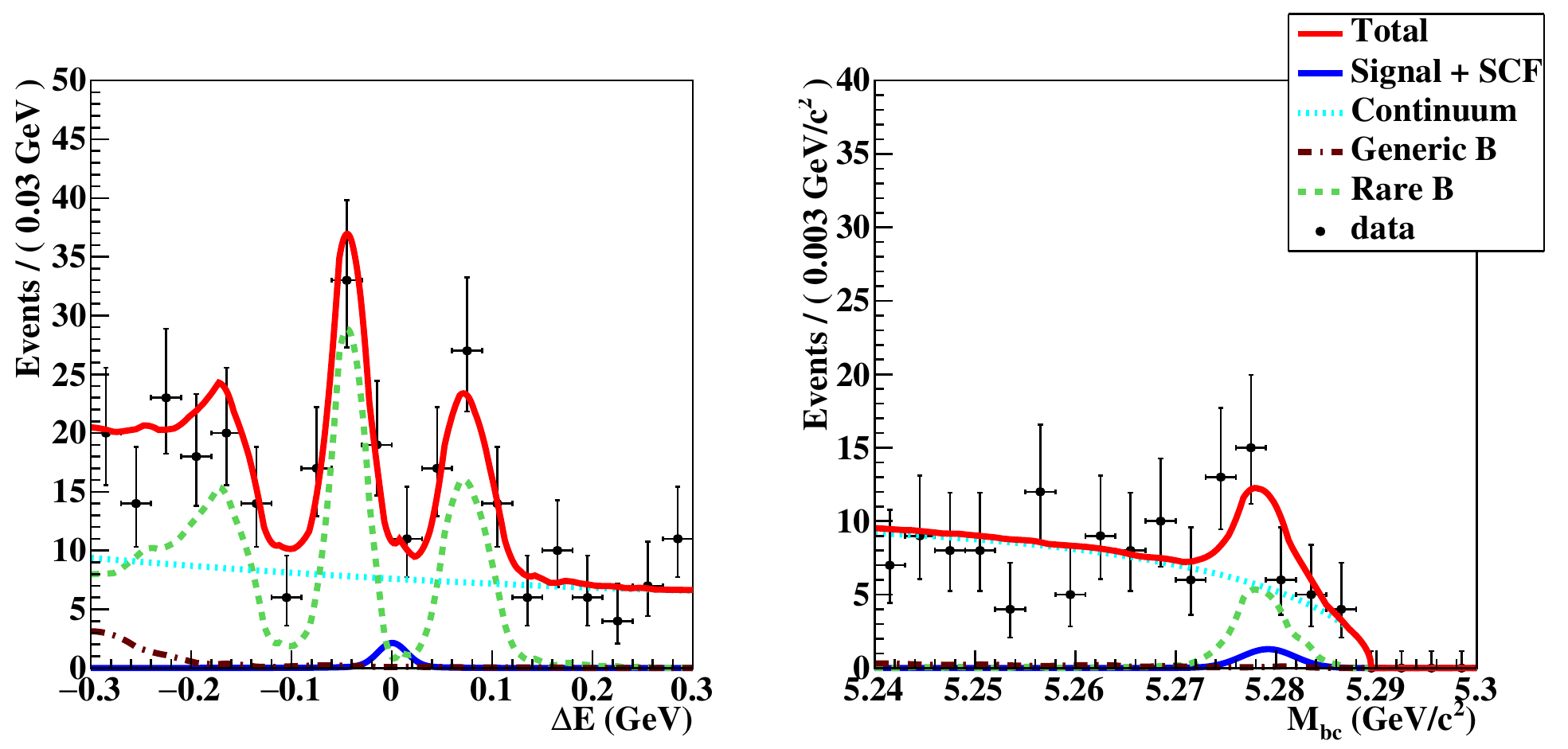}
 }
\caption{The projections of the $\mb$-$\de$ fit to data in the first $\mkk$ bin. 
Points with error bars are the data, the red line is the fit result, the blue line is the sum of the signal and the self cross feed, 
the cyan dotted line is the continuum background, the brown dash-dotted line is the generic $B$ backgrounds, 
and the green dashed line is the sum of the rare $B$ backgrounds. The projection on $\de$ is with the requirement of $5.275 < \mb < 5.2835$ GeV/$c^2$, while the projection on $\mb$ is with the requirement of $-0.03 < \de < 0.03$ GeV.
}
\label{fig:result_all}
\end{figure}

\begin{figure}
\includegraphics[width=0.23\textwidth]{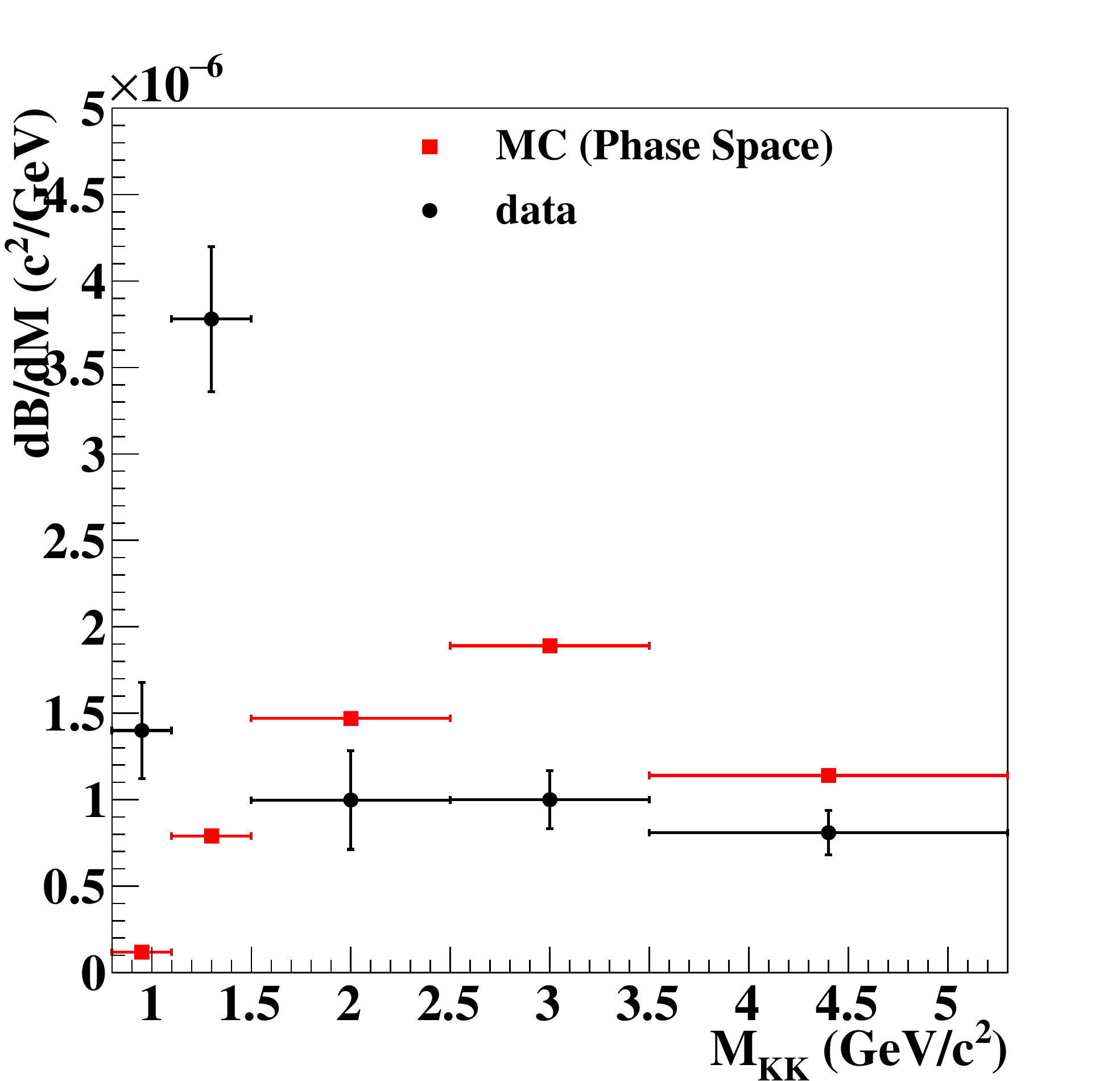} 
\includegraphics[width=0.23\textwidth]{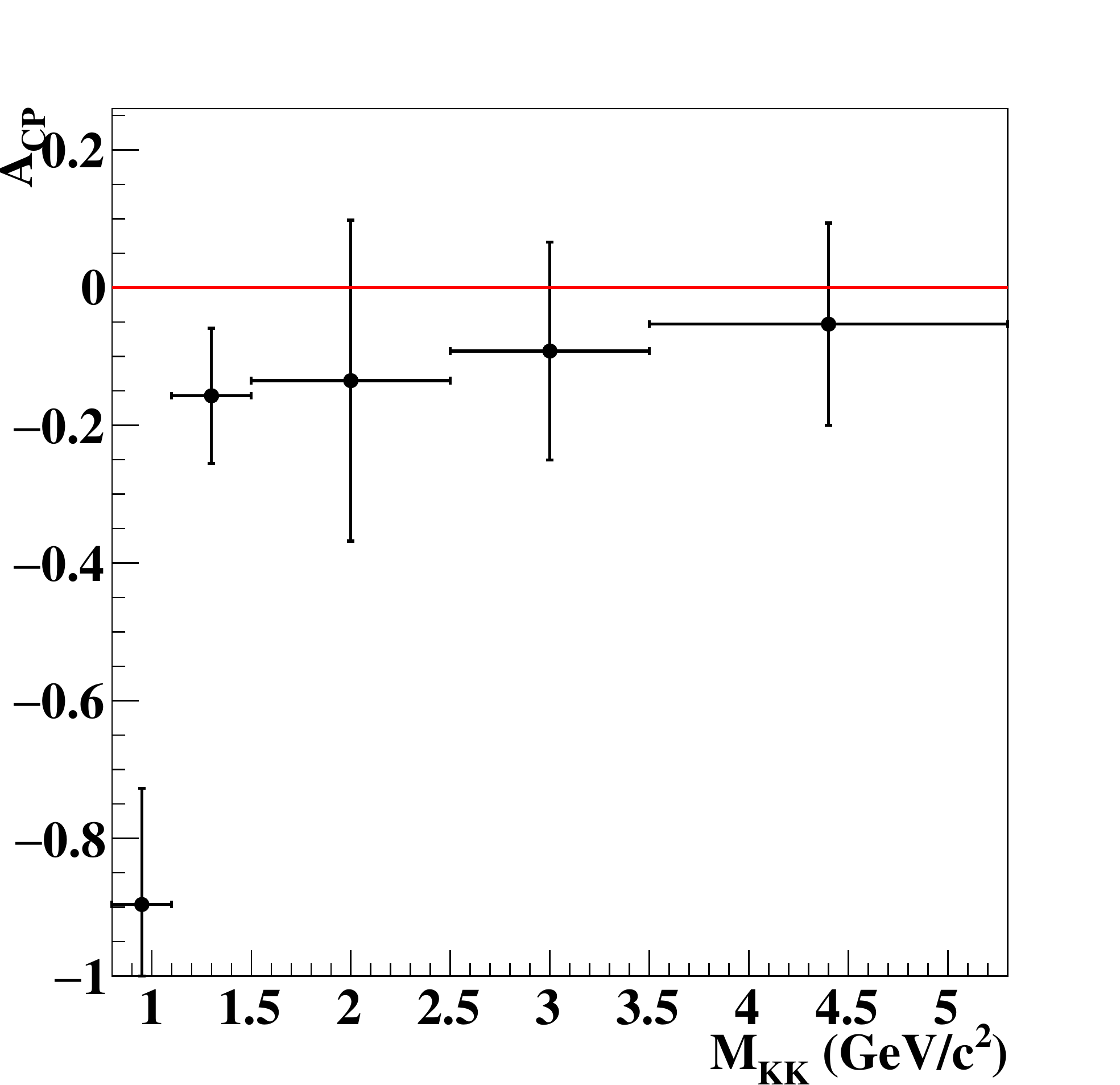} \\
\caption{
Differential branching fractions~(left) and measured $\acp$~(right) as a function of $\mkk$. Each point is obtained from a two-dimensional fit with systematic uncertainty included. Red squares with error bars in the left figure show the expected signal distribution in a three-body phase space MC. Note that the phase space hypothesis is rescaled to the total observed $\kkpi$ signal yield.}
\label{fig:binfit}
\end{figure}

As an example, Fig.~\ref{fig:result_all} shows the fit results in $B^{+}$ and $B^{-}$ samples of the first $\mkk$ bin in a signal-enhanced region. 
We use the efficiency and fitted yield in each bin to calculate the branching fraction:
\begin{equation}
\label{eq:bf}
\br=\frac{\Nsig}{\epsilon \times \cpid \times N_{\bb}}\mbox{,}
\end{equation}
where $N_{\bb}$, $\epsilon$, and $\cpid$, respectively, are the number of $\bb$ pairs ($772 \times 10^{6}$), the reconstruction efficiency, 
and the correction factor for particle identification ($94.2\%$) that accounts for possible data-MC difference.
We assume that charged and neutral $\bb$ pairs are produced equally at the $\Upsilon(4S)$ resonance.
Table~\ref{tab:binfit} lists the fitted yields, efficiencies, and measured $\acp$ in all such bins.
To determine the significance of our measurements, we use the convolution of the likelihood function with a Gaussian function of width equal to the additive systematic uncertainties that only affect the signal yield and $\acp$. The corresponding significance is given by $\sqrt{-2\ln(L_0/L_{\rm max})}$,
where $L_{\rm max}$ and $L_0$ are the likelihood values with and without the signal component, respectively.
Figure~\ref{fig:binfit} shows the results, where an excess and a large $\acp$ are seen in $\mkk < 1.5$ GeV/$c^{2}$, confirming the observations by $BABAR$ and LHCb.
We find strong evidence of a large $CP$ asymmetry of 
$-0.90\pm0.17\pm0.03$ with $4.8\sigma$ significance for $\mkk < 1.1$ GeV/$c^2$.
We integrate the differential branching fractions over the entire mass range to obtain an inclusive branching fraction:
\begin{equation}
\br(\kkpi)=(5.38 \pm 0.40\pm 0.35)\times 10^{-6}\mbox{,}
\end{equation}
where the quoted uncertainties are statistical and systematic, respectively.
The weighted average $\acp$ over the entire $\mkk$ region is
\begin{equation}
\acp = -0.170 \pm 0.073 \pm 0.017\mbox{,}
\end{equation}
where the $\acp$ value in each bin is weighted by the fitted yield divided by the detection efficiency in that bin.
The statistical uncertainties are independent among bins; thus, the term is a quadratic sum. 
For the systematic uncertainties, the contribution from the bin-correlated sources is a linear sum while the contribution from
the bin-uncorrelated sources is a quadratic sum.

\begin{table*}
\caption{Signal yield, efficiency, differential branching fraction, and $\acp$ for individual $\mkk$ bins. 
The first uncertainties are statistical and the second systematic. }
\label{tab:binfit}
\begin{center}
\begin{tabular}{c|cccc}
\hline\hline
$\mkk$  & \multirow{2}{*}{$\Nsig$} & \multirow{2}{*}{Eff. ($\%$)} & \multirow{2}{*}{$d\BR/dM~(\times 10^{-7})$} & \multirow{2}{*}{$\acp$} \\
(GeV/$c^2$) & & \\
\hline
$0.8-1.1$ & $59.8 \pm 11.4 \pm 2.6 $ & $19.7$ & $14.0 \pm 2.7 \pm 0.8$ & $-0.90 \pm 0.17 \pm 0.04$ \\
$1.1-1.5$ & $212.4 \pm 21.3 \pm 6.7$ & $19.3$ & $37.8 \pm 3.8 \pm 1.9$ & $-0.16 \pm 0.10 \pm 0.01$ \\
$1.5-2.5$ & $113.5 \pm 26.7 \pm 18.6$ & $15.6$ & $10.0 \pm 2.3 \pm 1.7$ & $-0.15 \pm 0.23 \pm 0.03$ \\
$2.5-3.5$ & $110.1 \pm 17.6 \pm 4.9$ & $15.1$ & $10.0 \pm 1.6 \pm 0.6$ & $-0.09 \pm 0.16 \pm 0.01$ \\
$3.5-5.3$ & $172.6 \pm 25.7 \pm 7.4$ & $16.3$ & $8.1 \pm 1.2 \pm 0.5$ & $-0.05 \pm 0.15 \pm 0.01$ \\
\hline\hline
\end{tabular}
\end{center}
\end{table*}

Systematic uncertainties in the branching fraction are itemized in Table~\ref{tab:sysm}.
The uncertainty due to the total number of $\bb$ pairs is $1.37\%$. The uncertainty due to the charged-track reconstruction efficiency is estimated to be $0.35\%$ per track by using the partially reconstructed $D^{*+} \to D^{0}\pi^{+}$, $D^{0}\to \pi^{+}\pi^{-}\ks$ events. 
The uncertainty due to the $\rkpi$ requirements is determined by a control sample study of $D^{*+}\to D^{0}\pi^{+}$ with $D^{0}\to K^{+}\pi^{-}$. 
The uncertainties due to the continuum suppression selection criteria and the signal PDF shape are estimated using a control sample of $\control$ decays. The potential fit bias is evaluated by performing an ensemble test comprising $1000$ pseudoexperiments, where the signal component is taken from the signal MC sample, and the PDF shapes are used to generate the data for the other event categories. The observed $2.3\%$ bias is included in the systematic uncertainty calculation.
The uncertainty due to the continuum background PDF modeling is evaluated by varying the PDF  
parameters by $\pm 1\sigma$ of their statistical errors.
The uncertainty due to the data-MC difference is taken into account by using the fit model from the off resonance data, which is included in the background PDF modeling in Table~\ref{tab:sysm}.
For the $\bb$  background PDFs that are modeled by a two-dimensional 
smoothed histogram PDFs, the associated uncertainty is evaluated by changing the bin sizes. 
The uncertainty due to the fixed yields of rare combinatorial backgrounds is also evaluated by varying each fixed yield up or down by its statistical error. 
The uncertainty due to nonzero $\acp$ of rare peaking backgrounds is estimated by assuming the $\acp$ values to be higher or lower than the LHCb measured values by $1\sigma$ of the LHCb measurement uncertainty~\cite{lhcb_new}.
In the absence of knowledge of the distribution of the SCF background in $\mkk$, we use a conservative approach to evaluate the uncertainty by varying the fraction by $\pm 50\%$; the resulting deviation from the nominal value is included in the fixed yields in Table~\ref{tab:sysm}.

The $\acp$ systematic errors due to the fixed yields, background $\acp$ and the background PDF modeling are estimated with the same procedure as applied for the branching fraction. 
A possible detector bias due to tracking acceptance and $\rkpi$ is evaluated using the measured $\acp$ value from the off resonance data. We apply the same criteria as those for the signal except for the continuum-suppression requirement and calculate $\acp$ as in Eq.~(\ref{eq:acpdef}).
The $\acp$ value from off resonance data is $0.0024 \pm 0.0014$. The final $\acp$ result can be either corrected with this detector bias, or it can be applied as a systematic uncertainty.
For this result the central shift plus $1\sigma$ statistical error is included in the total systematic uncertainty for the $\acp$. 
A full list of systematic uncertainties in $\acp$ is shown in Table~\ref{tab:sysm}.

\begin{table*}[htpb]
\caption{\label{tab:sysm}Systematic uncertainties in the measured branching fraction and $\acp$ in the individual bins. The dagger ($\dagger$) indicates the 
$\mkk$ dependence of the uncertainty. The center dots ($\cdots$) indicate a value below $0.05\%$~($0.001$) in $\BR$~(\acp).}
\begin{center}
\begin{tabular}{lccccc}
\hline\hline
{Source} & \multicolumn{5}{c}{Relative uncertainties in ${\BR}$ ($\%$)} \\
$\mkk$(GeV/$c^2$) & $0.8-1.1$ & $1.1-1.5$ & $1.5-2.5$ & $2.5-3.5$ & $3.5-5.3$ \\
\hline
Number of $\bb$ pairs & \multicolumn{5}{c}{$1.37$} \\
Tracking & \multicolumn{5}{c}{$1.05$} \\
Particle identification & \multicolumn{5}{c}{$1.44$} \\
Continuum suppression & \multicolumn{5}{c}{$1.33$} \\
Signal PDF & \multicolumn{5}{c}{$1.77$} \\
Fit bias & \multicolumn{5}{c}{$2.30$} \\
Background PDF$^\dagger$ & $3.65$ & $2.15$ & $16.16$ & $3.77$ & $3.59$ \\
Fixed yields$^\dagger$ & $\cdots$ & $\cdots$ & $\cdots$ & 0.07 & $\cdots$ \\
Background $\acp$$^\dagger$ & $0.23$ & $0.28$ & $1.46$ & $0.80$ & $0.36$ \\
\hline\hline
\multicolumn{2}{c}{} \\
\hline\hline
Source & \multicolumn{5}{c}{Absolute uncertainties in $\acp$} \\
$\mkk$(GeV/$c^2$) & $0.8-1.1$ & $1.1-1.5$ & $1.5-2.5$ & $2.5-3.5$ & $3.5-5.3$ \\
\hline
Background PDF$^\dagger$ & 0.036 & 0.005 & 0.028 & 0.006 & 0.003 \\
Fixed yields$^\dagger$ & $\cdots$ & $\cdots$ & $\cdots$ & 0.002 & $\cdots$ \\
Background $\acp$$^\dagger$ & 0.015 & 0.004 & 0.009 & 0.005 & 0.002 \\
Detector bias &  \multicolumn{5}{c}{$0.004$} \\
\hline\hline
\end{tabular}
\end{center}
\end{table*}

In conclusion, we have reported the measured branching fraction and direct $CP$ asymmetry for the suppressed decay $\kkpi$ using the full $\Upsilon(4S)$ data sample collected with the Belle detector. 
We employ a two-dimensional fit to determine the signal yield and $\acp$ as a function of $\mkk$.
We confirm the excess and local $\acp$ in the low $\mkk$ region reported by LHCb, and quantify the differential branching fraction in each $\kk$ invariant 
mass bin. We find a $4.8\sigma$ evidence for a negative $CP$ asymmetry in the region $\mkk < 1.1$ GeV/$c^{2}$. 
Our measured inclusive branching fraction and direct $CP$ asymmetry are 
$\BR(\kkpi) = (5.38\pm0.40\pm0.35)\times 10^{-6}$ and $\acp = -0.170\pm0.073\pm0.017$, respectively. 
The measurement challenges the conventional description of direct $CP$ violation since it requires large contributions to separate weak tree and strong penguin amplitudes in the same small region of phase space in order to simultaneously enhance both the yield and provide the cancellation required for such a large $CP$ effect. So, for example, if the enhancement were due to a large final state resonance in a strong penguin diagram, there would have to be an accompanying tree-level process of the same magnitude and opposite phase to provide the almost complete cancellation observed in the measurement.
To understand the origin of the low-mass dynamics, a full Dalitz analysis from experiments with a sizeable data set, such as LHCb and Belle II, will be needed in the future.

We thank the KEKB group for the excellent operation of the
accelerator; the KEK cryogenics group for the efficient
operation of the solenoid; and the KEK computer group,
the National Institute of Informatics, and the 
PNNL/EMSL computing group for valuable computing
and SINET5 network support.  We acknowledge support from
the Ministry of Education, Culture, Sports, Science, and
Technology (MEXT) of Japan, the Japan Society for the 
Promotion of Science (JSPS), and the Tau-Lepton Physics 
Research Center of Nagoya University; 
the Australian Research Council;
Austrian Science Fund under Grant No.~P 26794-N20;
the National Natural Science Foundation of China under Contracts 
No.~10575109, No.~10775142, No.~10875115, No.~11175187, No.~11475187, 
No.~11521505 and No.~11575017;
the Chinese Academy of Science Center for Excellence in Particle Physics; 
the Ministry of Education, Youth and Sports of the Czech
Republic under Contract No.~LTT17020;
the Carl Zeiss Foundation, the Deutsche Forschungsgemeinschaft, the
Excellence Cluster Universe, and the VolkswagenStiftung;
the Department of Science and Technology of India; 
the Istituto Nazionale di Fisica Nucleare of Italy; 
the WCU program of the Ministry of Education, National Research Foundation (NRF)
of Korea Grants No.~2011-0029457, No.~2012-0008143,
No.~2014R1A2A2A01005286,
No.~2014R1A2A2A01002734, No.~2015R1A2A2A01003280,
No.~2015H1A2A1033649, No.~2016R1D1A1B01010135, No.~2016K1A3A7A09005603, No.~2016K1A3A7A09005604, No.~2016R1D1A1B02012900,
No.~2016K1A3A7A09005606, No.~NRF-2013K1A3A7A06056592;
the Brain Korea 21-Plus program, Radiation Science Research Institute, Foreign Large-size Research Facility Application Supporting project and the Global Science Experimental Data Hub Center of the Korea Institute of Science and Technology Information;
the Polish Ministry of Science and Higher Education and 
the National Science Center;
the Ministry of Education and Science of the Russian Federation and
the Russian Foundation for Basic Research;
the Slovenian Research Agency;
Ikerbasque, Basque Foundation for Science and
MINECO (Juan de la Cierva), Spain;
the Swiss National Science Foundation; 
the Ministry of Education and the Ministry of Science and Technology of Taiwan;
and the U.S.\ Department of Energy and the National Science Foundation.


\begin{thebibliography}{99}
\bibitem{BaBar_kkpi} B.~Aubert {\it et al.} ($BABAR$ Collaboration), Phys. Rev. Lett. {\bf 99}, 221801 (2007).
\bibitem{lhcb_old} R.~Aaij {\it et al.} (LHCb Collaboration), Phys. Rev. Lett. {\bf 112}, 011801 (2014).
\bibitem{lhcb_new} R.~Aaij {\it et al.} (LHCb Collaboration), Phys. Rev. D {\bf 90}, 112004 (2014).
\bibitem{Bhattacharya2013} B.~Bhattacharya, M. Gronau, and J. L. Rosner, Phys. Lett. B {\bf 726}, 337 (2013).
\bibitem{Bediaga2014} I.~Bediaga, T.~Frederico, and O.~Louren\ifmmode \mbox{\c{c}}\else \c{c}\fi{}o, Phys. Rev. D {\bf 89}, 094013 (2014).
\bibitem{kekb} S.~Kurokawa and E.~Kikutani, Nucl. Instrum. Methods Phys. Res., Sect. A {\bf 499}, 1 (2003), and other papers included in this volume. T.~Abe {\it et al.}, Prog. Theor. Exp. Phys. {\bf 2013}, 03A001 (2013), and references therein.
\bibitem{belled} A.~Abashian {\it et al.} (Belle Collaboration), Nucl. Instrum. Meth. A {\bf 479}, 117 (2002); also see detector section in J.~Brodzicka {\it et al.}, Prog. Theor. Exp. Phys. {\bf 2012}, 04D001 (2012).
\bibitem{bellesvd2} Z.~Natkaniec {\it et al.} (Belle SVD2 Group), Nucl. Instrum. Methods Phys. Res., Sect. A {\bf 560}, 1 (2006).
\bibitem{geant} R.~Brun {\it et al.}, {\textsc GEANT} 3.21, CERN Report No. DD/EE/84-1, 1987.
\bibitem{evtgen} D.~J. Lange, Nucl. Instrum. Methods Phys. Res., Sect. A {\bf 462}, 152 (2001).
\bibitem{neubay} M.~Feindt and U. Kerzel, Nucl. Instrum. Methods Phys. Res., Sect. A {\bf 559}, 190 (2006).
\bibitem{ksfw} G.~C.~Fox and S.~Wolfram, Phys. Rev. Lett. {\bf 41}, 1581 (1978); The modified moments used in this paper are described in S.~H.~Lee {\it et al.} (Belle Collaboration), Phys. Rev. Lett. {\bf 91}, 261801 (2003).
\bibitem{thrust} S.~Brandt, C.~Peyrou, R.~Sosnowski, and A.~Wroblewski, Phys. Lett. {\bf 12}, 57 (1964).
\bibitem{btagging} H.~Kakuno {\it et al.} (Belle Collaboration), Nucl. Instrum. Methods Phys. Res., Sect. A {\bf 533}, 516 (2004).
\bibitem{argus} H.~Albrecht {\it et al.} (ARGUS Collaboration), Phys. Lett. B {\bf 241}, 278 (1990).
\bibitem{pdg2016} C.~Patrignani {\it et al.} (Particle Data Group), Chin. Phys. C, {\bf40}, 100001 (2016).

\end{thebibliography}
\end{document}